# Reciprocity Deficits

Observing AI in the street with everyday publics

Alex Taylor[1], Noortje Marres, Mercedes Bunz, Thao Phan, Maya Indira Ganesh, Dominique Barron, Yasmine Boudiaf, Rachel Coldicutt, Iain Emsley, Beatrice Gobbo, Louise Hickman, Manu Luksch, Bettina Nissen, Mukul Patel, Luis Soares

## ABSTRACT

The street has emerged as a primary site where everyday publics are confronted with AI as an infrastructural phenomenon, as machine learning-based systems are now commonly deployed in this setting in the form of automated cars, facial recognition, smart billboards and the like. While these deployments of AI in the street have attracted significant media attention and public controversy in recent years, the presence of AI in the street often remains inscrutable, and many everyday publics are unaware of it. In this paper, we explore the challenges and possibilities of everyday public engagement with AI in the situated environment of city streets under these paradoxical conditions. Combining perspectives and approaches from social and cultural studies of AI, Design Research and Science and Technology Studies (STS), we explore the affordances of the street as a site for 'material participation' in AI through design-based interventions: the creation of 'everyday AI observatories.' We narrate and reflect on our participatory observations of AI in five city streets in the UK and Australia and highlight a set of tensions that emerged from them: 1) the framing of the street as a transactional environment, 2) the designed invisibility of AI and its publics in the street 3) the stratification of street environments through statistical governance. Based on this discussion and drawing on Jane Jacobs' notion of "eyes on the street," we put forward the relational notion of "reciprocity deficits" between AI infrastructures and everyday publics in the street. The conclusion reflects on the consequences of this form of social invisibility of AI for situated engagement with AI by everyday publics in the street and for public trust in urban governance.

## 1 Introduction

How to co-exist with AI in the street? Is it desirable or even possible? Over the last ten years or so, the street has emerged as a primary site where everyday publics encounter AI, in the form of self-driving vehicles, facial recognition cameras, smart traffic lights and smart lampposts, AI-enabled billboards, and in some suburban contexts, delivery drones. Initially taking the form of trials and pilots conducted by private companies and public sector organisations including the police and local transport authorities, AI-based technologies are now deployed routinely in many streets across the United Kingdom and elsewhere. Several of these deployments have been high profile, as in the case of government-funded street trials with autonomous vehicles in the late 2010s [26], and some have proven highly controversial, as the use of facial recognition by the South Wales police was deemed unlawful by the UK High Court in 2020, a verdict that was followed by efforts including by UK parliamentarians to ban facial recognition. At the same time, however, the deployment of AI-based technologies in the street has continued and indeed expanded in the United Kingdom as elsewhere. In many cases, passers-by, residents and workers in the streets in which AI is being deployed remain unaware of its presence.

In this paper, we tackle the paradoxical and problematic nature of AI in the street by narrating and reflecting on five everyday AI observatories realised in four cities across the UK and one in Australia in the summer of 2024. Taking the form of diverse design-based participatory interventions, we created these street-level observatories of AI in urban settings with two aims: 1) to complicate and counter overidealized accounts of AI in society by rendering AI explorable as a messy, everyday reality and 2) to invite everyday publics to explore with us often invisible smart urban infrastructure in their lived environments. However, in realising our everyday observatories of AI in city streets with residents and visitors, we encountered a further challenge that we had not fully anticipated, namely the stubborn inaccessibility of AI as an infrastructural phenomenon in the street.

---

[1] Contact Author – University of Edinburgh, email: alex.taylor@ed.ac.uk

In all streets where we conducted our observatories AI was an undeniable presence, in the form of automated vehicles, intelligent traffic management technologies and smart billboards. Yet we found that in many if not most cases, the presence of AI remained hard to grasp, and proved infrastructurally inscrutable. Our observatories show that this inscrutability of AI in lived environments does not just derive from the often-noted 'opacity' of machine learning-based technologies. It is equally associated with a social form of invisibility, as participants expressed a sense of feeling excluded and 'left out of the loop' by AI and data infrastructures in the street. To denote this situation, we put forward the term 'reciprocity deficit. Drawing on the well-known urbanist and writer Jane Jacobs' ideas of reciprocity in casual, ordinary living on the street, we show how computational and algorithmic modes of addressing the world are often felt to ignore and be detached from life in the street and as such become part of a continued erosion of webs of public respect and trust.

The paper is structured as follows. We begin by introducing the street as a setting for everyday public engagement and 'material participation' in AI and its problems, drawing on recent social and cultural studies of AI, design research and Science and Technology Studies. Turning to methodology, we introduce our everyday observatories of AI, showing how combining methods from art-based research and participatory design with data mapping approaches developed in STS allows us to explore the infrastructural phenomenon of AI in the street with everyday publics. We go on to discuss observations from across the five observatories and develop our analysis of the reciprocity deficits of urban AI. The conclusion sums up what we consider the main contribution of this paper, namely, to evaluate AI in the street from the standpoint of its operations and effects on reciprocity relations in the street.

## 2  The street as a critical and challenging site for public participation in AI

Over the last ten years, concerns about harms, risks, and accountability of AI systems have transformed AI systems, digital interfaces and machine learning models into notable objects of public engagement. In the UK, the street stands out as an especially notable setting for such public engagement with AI systems, as trials of AI-based technologies like automated vehicles and facial recognition have been conducted in this setting since 2015 with significant government support, attracting high levels of media attention and, in the case of facial recognition, eliciting activist mobilisation, public controversy and regulatory intervention by the

UK's data protection watchdog, the ICO, in 2019 (for a discussion see [14, 14, 39].

Activists, experts, journalists and citizens have identified and demonstrated several significant challenges raised by AI in the street that both directly affect everyday publics and impact the deeper institutional underpinnings of the street as a public space. These include 1) increased surveillance and monitoring of people and life in the street, especially in city centres and more deprived urban neighbourhoods, and the biased nature of AI-based surveillance [16]; 2) lack of democratic legitimacy and public consent as the deployment of AI-based technologies in the street often remains undisclosed to affected publics, in the UK especially in the early trial phases (2016-2020). 3) the delegation of public sector responsibilities, such as suspect identification for law enforcement purposes and traffic management by transport authorities, to private companies [6]; 4) the growing disjuncture between high levels of state investment in privately managed technological infrastructures in the street alongside sustained lack of investment in community-based facilities in these same environments ("liveable neighbourhoods") under conditions of public sector cuts and austerity [26].

Partly in response to the demonstration of such political, ethical, socio-technical and regulatory challenges raised by the deployment of AI in the street, there have been various efforts in support of public engagement with AI innovation in the street in the UK, in the form of public demo's of technologies like self-driving cars [31], participatory audits of facial recognition technologies [16] and design-led user workshops on automated mobility futures hosted in public settings such as the transport museums in Coventry and London [26]. However, efforts to design more enduring forms of public transparency, accountability and engagement into AI systems in the street have been less forthcoming and have proven difficult for several reasons.

With a particular focus on cities, Gupta and Loukisssas [17, 18] draw attention to key weaknesses in efforts to implement frameworks of *explainable AI* (XAI)—where fairer AI systems are sought through publicly legible explanations of algorithmic operations – in urban environments. They argue that AI in cities has uneven and sometimes discriminatory spatial effects that remain opaque and are not accounted for in established approaches to XAI. As they express it:

> "The overly simplistic mathematical representations that algorithms rely on are bound to amplify some features of cities and flatten others. These representational distortions can create the kinds of invisible inequalities and incremental injustices that produce large-scale societal effects, such as gentrification and segregation..." [p. 5, 17]

Emphasising how the introduction of AI in the street can amplify existing socio-economic dynamics, Gupta and Loukisssas propose situated and spatial representations of AI algorithms that go beyond the 'mathematical' with the aim of making visible encounters with AI that are grounded in lived experiences. This work suggests that the understanding and facilitation of public engagement with AI in the street stands much to learn from design-led approaches that have been developed in the area of digital urban civics over the last ten years or so. This work deploys situated methods to facilitate civic engagement in urban spaces, building capacities for 'material participation' [11, 4], in work that centres the role of more-than-human settings such as cycling lanes and street crossings as devices to engage different groupings and communities including cyclists [1, 30] and disability campaigners [23] in addressing social and environmental problems in urban environments.

While there is thus a significant body of work demonstrating the contemporary relevance of the street as a setting for situated, material engagement by everyday publics in technological and ecologically challengedsocieties, the deployment of AI in urban environments also raises particular challenges in this regard. This notably includes the invisibility AI in the street. Social studies of AI in everyday life have long emphasised the opaque nature of these systems [6, 8]. As we will go on to discuss in more detail below, in the street AI frequently presents as an inscrutable phenomenon, of which the very existence can easily be called into question: Is AI really being deployed in this street? How do you know? As we will discuss in more detail in what follows, the lived experience of AI in the street often centres on its inaccessibility and tenuous relation to the street as a lived environment, resulting in its articulation as a fundamentally withdrawn and indifferent presence in the street, as "machines that are literally over our heads" (Coventry observatory). What forms of everyday public engagement with AI are relevant, possible or indeed desirable under these conditions of the "absent presence" of AI in the street?

Indeed, there are extensive critiques of lay and public participation in AI as set out by Sloane et al [37], Corbett et al. [10], Gourlet et al [15]. This literature argues that machine learning-based infrastructures are always already participatory, insofar as their functioning depends on the capture of large masses of user-generated data and continuous uptake of these systems by publics in and as everyday life. For this reason, it has been argued that contemporary AI-based systems configure the relation between technology and the public in inherently extractive ways. Using the conceptual repertoire of the social studies of technology, we can say that the introduction and presence of machine-learning based systems in everyday living environments like the street amounts to the "resourcification" of participation [22], whereby publics through their everyday actions enable the functioning and valuation of these infrastructural systems, even as they often remain unaware of their operations.

The configuration of everyday urban environments as sites for frequently extractive and non-voluntary forms of participation, as a consequence of the introduction of AI-based systems into city streets, raises new challenges for the facilitation of material participation with AI in urban settings. Not only is AI innovation often opaque and invisible from the standpoint of everyday life in the street. We must also ask anew whether and how situated and materially aware forms of participation in this everyday environment can really provide opportunities to redistribute power and agency between publics and other, often more powerful, entities, such as the state, industries or infrastructures? While the infrastructuring of participatory actions by state and industry has a long history, with form filling by bureaucratic subjects to citizen involvement in infrastructure management (such as flood control), the ubiquitous extraction and direct valorisation of user data and actions within contemporary computational, increasingly AI-enabled - and enabling - architectures in society feels unprecedented.

## 4 Street-level observatories of everyday AI

The introduction of AI in the street today is occurring in a distinctive context, one in which infrastructure creation in cities is closely associated with top-down government investment in commercial solutions for traffic management and safety in a context of austerity. The notion of 'material participation' nevertheless remains relevant to everyday engagement with AI in the street insofar as it highlights the 'underdeterminacy' of participation : just as in the case of cycling lanes, technologies such as smart traffic management systems do not just 'enrol' everyday subjects in urban governance and impose a "programme of action" on them via socio-technical systems. Such material settings and devices may equally open up opportunities for the negotiation, contestation and problematization of technological governance through everyday material practices [19].

However, efforts to create occasions for material participation in AI in the street face a distinctive challenge insofar as "AI" is not transparently available as an object—or even a focal point—of public engagement in the lived environments in the street: not only are residents in and visitors to streets in which AI is deployed often unaware of its presence, it is far from self-evident whether a given vehicle, smart billboard, networked lamp post, sensor, or camera really can be defined as "AI-enabled." The disputed presence and partial manifestation of AI in the street requires us to place special emphasis on the observation of AI in the street: for everyday engagement with AI to be possible in the street, the phenomenon, AI, must be actively identified, made visible, and/or elicited in this setting. It was to experiment with this that we devised a makeshift methodology in the summer of 2024, which we called a "street-level observatory of everyday AI."

City streets have been designated as sites for research in a variety of projects across design and social studies, many of which take advantage of the affordances of the street as a relatively open, inherently participatory and/or public environment. The *Tidy Street* project [3], for instance, involved a residential street in monitoring its energy consumption, and used public, low-fi visualisations (chalked on the tarmac) to prompt changes in energy use. A longer-term engagement with *Tension Road* centred on the role data might play for residents and involved a variety of interventions to experiment with situated forms of participation [24, 40]. *Streets for*

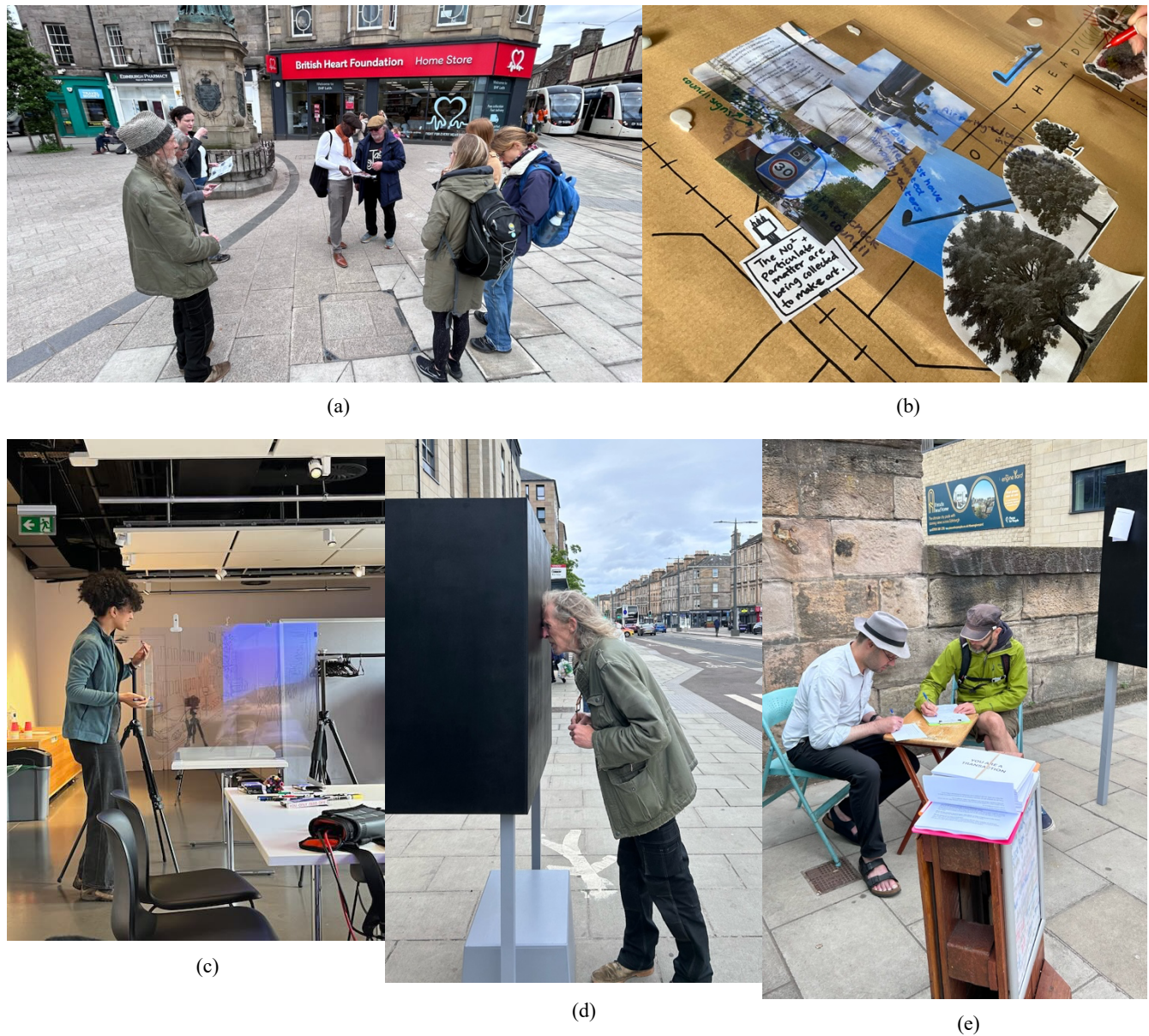

**Figure 1.** (a) Data walk in Edinburgh with residents/users of street, (b) participant's output from mapping exercise in Coventry, (c) setup of projected map for participants to annotate in London workshop, and Edinburgh "black box (d) and passersby responding to card prompts (e).

*People* [34], a more geographically dispersed project, involved a number of streets to investigate placemaking and children's roles in urban design proposals. These projects used the street not as a site to produce generalisable claims about the populations that pass through them, but, to the contrary, to use the situated setting of the street to convene situated actors to articulate and elicit awareness of the specific infrastructural underpinnings of everyday life in technological societies.

The research we present here —based on five everyday AI observatories—was motivated by similar aims. From the outset, we defined an AI observatory in a loose and open way. Our aim was to develop AI observation into a situated method for engagement and co-inquiry with everyday publics by anchoring it in the street. We found inspiration in site specific interventions by artists and designers in city streets, such as Polak and Van Bekkum's [35] *Spiral Sunrise* and Niederer et al.'s [33] street-level city analytics: we then took up place-based digital and embodied methods of participatory research, such as data walks and locative data mapping, to curate occasions for co-observation of the situated effects of AI innovation in the everyday environment of the street.

Reciprocity deficits

To realise our observatories, AI in the street brought together five interdisciplinary research teams composed of social and cultural researchers, designers and artists based in Universities across the United Kingdom as well as in Melbourne (Australia). Each of these teams identified a particular street in near-by cities where AI is or was recently deployed (See Table 1 for overview of the observatories).

The selected streets can be distinguished both in terms of the type of street and the nature of AI deployment: the observatory in Coventry was conducted in a residential commuter street that forms part of the Intelligent Mobility Testbed mentioned in the introduction, and which had been modified to facilitate automated vehicle testing. In Cambridge, AI in the street researchers conducted a Data Access Walk in the historic city centre: this walk was focused on the role of AI in accessibility issues and involved participants from local government responsible for the development of smart mobility infrastructure for the city. In London, artist-researchers selected two urban thoroughfares with a high density of different actual and possible AI applications for participatory data mapping and annotation. The observatory in Logan extends an Australian Research Council funded project investigating the social and cultural impacts of AI trials in Australia, with a focus on the trial and testing of commercial drone delivery systems. Edinburgh narrowed its engagement to a long, part-residential-part-commercial, street joining two historically distinct neighbourhoods, and involved residents and users of the street during controversial changes to transport infrastructure and housing, and accompanying tensions over gentrification.

As to the chosen methods, we prioritised the fit between team expertise, the features of the location, and the site-specific operationalisation of AI in each street. As a consequence, the different observatories implemented diverse methods, over the eight-month duration of the project (between February to September, 2024), though most made use of walking and diagramming. Each observatory worked with small groups of recruited and local participants, and research activities included data and sensing walks [31, 36] (Fig. 1a), different diagramming and mapping exercises (Fig. 1b/c), and individual interviews and group workshops. As part of the Edinburgh observatory, there was also a street intervention in which a large "black box" (containing a screen displaying text prompts informed from earlier data walks) was installed on the street to both engage passersby (Fig. 1d) and invite written responses on printed cards (Fig. 1e). All observatories sought out and worked with local groups and stakeholders, prioritising collaboration with arts and design agencies.

Our analysis involved multiple readings and discussions of materials, annotations and recordings collected from the observatories. Researchers at each observatory analysed their own observatory results guided by prompts devised by the project lead team and during several synthesis meetings we read across the results from different locations. Different stages of reporting and analysis were documented and shared throughout: in-person and remote brainstorming sessions over the course of the project;

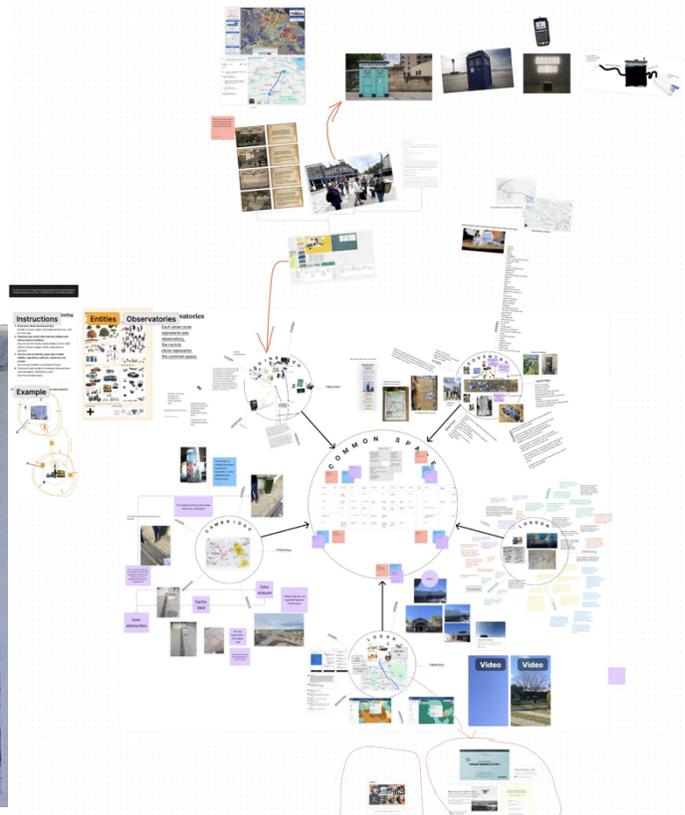

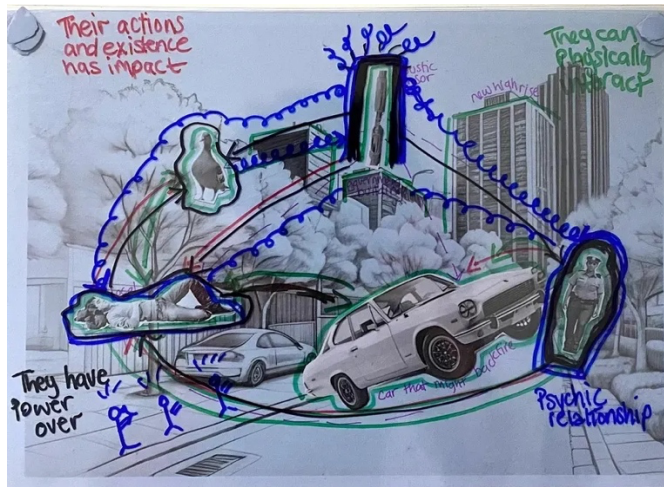

(a)            (b)

**Figure 2.** (a) Example of project team's diagram representing encounters of AI in the street and (b) diagram of prompt responses from all five observatories (details purposefully omitted and shown for illustrative purposes).

| Location | Participants/Partners | Methods | Data |
|---|---|---|---|
| City centre, Cambridge | Local government representatives working in city planning and technology infrastructures. | Accessibility data walks; post-walk group discussion. | Audio/video recordings captured by researchers. |
| Intelligent vehicle test site, Holyhead Road, Coventry | Local residents; local arts company. | Group "sensing walk"; post-walk diagramming workshop. | Images captured by researchers; maps/diagrams produced by participants. |
| Leith Walk, Edinburgh | Local residents and users of street; public service representatives; creative sector organisations; public sector partnership advising national govt on AI. | Multiple individual and group data walks; post-walk discussions; street intervention and card-based prompts targeting passersby. | Images and audio captured by researchers; Whatsapp channel with participant images and 'tags'; maps produced by participants; written responses to card-based prompts. |
| Drone delivery trial area, Logan | Local residents and small business workers/owners; documentary film maker. | Crowdsourcing stories of everyday interactions with drones; documentary film-making | Social media content; video/audio recordings of Interviews. |
| South and East London | Local residents; local artists. | Mapping/ diagramming workshops at three sites | Images and audio/video recordings captured by researchers; maps/diagrams produced by participants. |

**Table 1.** Overview of Everyday AI Observatories including their locations, the institutions where research teams were based, participants and partners involved in the research, methods used and data captured.

collective visual diagramming exercises (Fig. 2a); and the synthesis of visual and textual responses to prompts from each of the observatory teams supported by Figma (Fig. 2b). These modes of data capture, clustering and analysis resulted in a rich array of materials and insights that continue to be iteratively worked through.

The observations that we report on in this paper were identified as notable local issues and in several cases they speak to common themes arising from our shared process of analysis. Through group discussions and collaborative writing, these issues and problematics were put in conversation with related and contemporary theorising in social and cultural studies of AI, urban and infrastructure studies, participatory- and city-related literatures, critical data studies and platform urbanism to deepen our analysis and to help situate the work against broader critical arguments.

All the project's observatories were subject to research ethics review and approved by respective institutional review panels at the universities of ▮▮▮▮▮▮▮▮▮▮▮▮▮▮▮▮▮▮▮▮▮▮▮▮▮▮. Pseudonyms have been adopted in this paper to preserve the anonymity of the participants. Where participants are identifiable in photographs, permission has been given.

# 5 Observations and shared insights

We report on the observations and shared insights that emerged from our collective research by centring three ways of framing AI in the street that arose across the observatories: the street as a transactional site, the designed invisibility of AI and the street as a site for governance. These observations inform our notion of the reciprocity deficits of AI in the street which we will present in the conclusion.

## 5.1 The Transactional Street

In two of our observatories, in particular, in Edinburgh and Logan, there was a strong sense that data technologies, computation and AI were being implemented in the street as part of broader initiatives to create transactional infrastructures in society. These logics were often defined by participants in terms of a mundane calculus: access to products, services and infrastructure was provided in exchange for personal data and present or future market share. Examples that were highlighted included the use of loyalty cards, mass transit payment tracking, the consent required to use commercial, supermarket apps; and pedestrian capture and counting via street cameras. This is also to say that "AI" in the street was often framed as a component that enabled expansion of wider tech infrastructures.

For example, after a group data walk in Edinburgh, participants discussed a large electronic billboard placed along the city's Leith Walk and the installed camera pointing towards it (Fig. 3). Making connections to credit card companies and payment card readers in their unfolding discussions, the participants began to narrate the presence of this additional networked data infrastructure in terms of wider transactional logics:

**Ned**: "That camera just tells the advertising company in London when the screen goes down. So I often feel that some of the advertising has nothing to do with Edinburgh."

**Suzanne**: "Lots of things advertised … are disingenuous. They're not really benefiting the people. Like credit card companies – things that are causing more problems, really. And then next to that there are single traders that are using card readers."

The larger technological and financial systems are cast here as self-interested, removed from local concerns. This was also captured by passers-by in their written responses to the card prompts during the Leith Walk design intervention (Fig. 4). The broad sense was that a transactional logic prevails in the installation of data and AI-enabling infrastructure in the street. Structures and processes— from *in situ* data collection to real-time monitoring and aggregate analytics—are felt to prioritise business and commercial interests over the everyday people whose lives are intertwined with the street.

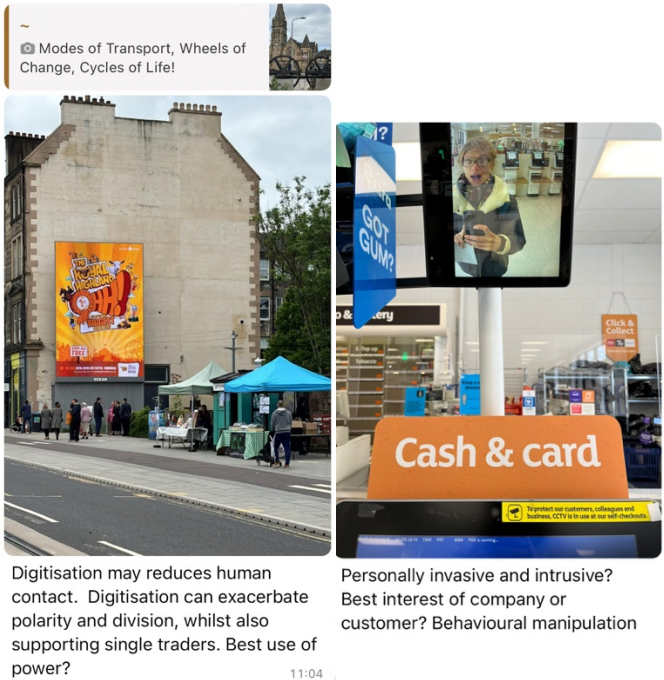

Digitisation may reduces human contact. Digitisation can exacerbate polarity and division, whilst also supporting single traders. Best use of power?

Personally invasive and intrusive? Best interest of company or customer? Behavioural manipulation

**Figure 3.** WhatsApp posts shared between Edinburgh data walk group.

A similar sentiment was expressed in Logan where residents described their frustration at the prioritisation of commercial interests over local needs. Although street residents and local shop owners had initially felt positive about the introduction of commercial drone delivery trials in their community, this quickly changed when the trial period ceased and the company administering the service, Wing Aviation (an Alphabet subsidiary), began to pivot their operations towards larger corporate partners; For local residents, the initial presence of Wing in Logan was a sign of positive change and overdue economic investment in a relatively disadvantaged city. Residents were impressed with the level of innovation and the quality of service suddenly made available to them. However, public sentiment shifted once the trial phase was complete and the company began to recoup costs through exclusive partnerships with major chains at the expense of smaller, local businesses.

The extractive nature of the data and computational infrastructures introduced under the rubric of trialling autonomous technologies, driven by commercial logics and concentrated wealth expansion, has been well documented (e.g., [5, 41, 43]). Here, at street-level, it becomes clear how this extraction is both profoundly ambivalent in its opening up of a space of transaction, while at the same time creating the context for the reassertion of familiar divides between those who live and work on the street and those benefitting from AI innovation in these living environments.

For Edinburgh and Coventry participants, questions arose over the different motivations and interests bound up with computationally enabled infrastructure in the streets. On a data walk in Edinburgh, Suzanne spoke of the cameras and matters of surveillance:

"It's one thing to have a camera there, like, as a woman right. You're sitting at a bus stop. But if you were attacked… no one's looking at that camera on a regular basis saying I'm going to send help right now. That type of action does not happen. You're going to look at that camera after the incident happened, after the fact…"

Participants thus observed a sort of detachment from happenings on the street enabled by networked data infrastructures and the services they enable, like CCTV-enabled crime detection: noting an implied indifference, a disjuncture, or what might be thought of as a "disentanglement" [42], between the different scales and time-space of embodied interaction in the street —the situations of unsafety that render 'AI' analytics situationally relevant —and the way data and AI infrastructures and processes are designed to operate as surveillance technology. Suzanne's comment indicates she is aware the cameras aren't there to trigger a human response, they are there for another purpose, perhaps to aggregate data on criminality and justify policing in specific areas. Whatever the case, the systems are understood to be designed for purposes that are detached or disentangled from what happens in situ.

In Logan, this line of reasoning is expanded. Reflecting on what are seen as the business motives behind the autonomous drone delivery trial, a local resident spells out the misalignment of interests:

"They didn't come and ask the people in Logan if they wanted it to happen. I'd prefer more infrastructure, more to public transport, more to the upkeep of parks and recreation, More advancement, you know, in the area, rather than drones. It doesn't benefit, you know, the people that are living here. There's a housing crisis."

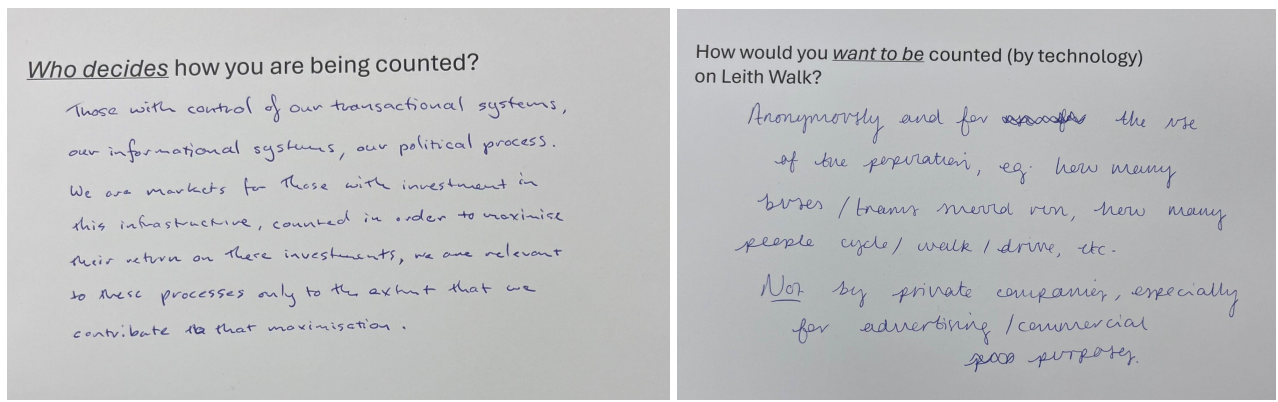

"Those with control of our transactional systems, our informational system, our political process. We are markets for those with investments in this infrastructure, controlled in order to maximise their return on their investment, we are relevant to these processes only to the extent that we contribute to that maximisation."

"Anonymously and for the use of the population, e.g., how many buses/trams would run, how many people cycle/walk/ drive, etc. Not by private companies, especially for advertising/commercial purposes"

**Figure 4.** Written responses to the card prompts during the Leith Walk design intervention

You know, look after the people don't... that doesn't benefit anybody."

The logic of innovation that finds a material form in trials with AI enabled drones in their city is presented as in tension but ultimately indifferent to local interests. AI systems and those deploying them show, in Logan, at least, a highly tactical - transactional - interest in what matters to people on the ground. While many interviewees expressed their excitement about big tech choosing Logan as their trial site, and willingly participated in the drone delivery trial by ordering coffee and food for home delivery by drone, several of them equally concluded that the benefits in the end go somewhere else, not to the people of Logan.

The introduction of data and AI-enabling technologies and infrastructures in the street thus seem to create or aggravate asymmetries in the embodied socio-economic relations that make up life in this setting. For those living and working on the street, the wish is to see benefits of AI innovation for the local community: to see AI making a difference to one's safety or for ordinary life, to transport, parks, homes, safety. etc. However, the overriding sense is that data and AI systems are being designed to operate at a scale that is detached from the specificity of the street as a lived environment, from its unique geographies, histories, social forms and vulnerabilities. There is a recognition that data's and AI's operational scales supersede or bracket situated and local contingencies; fractures come to light between the street-based operations that data and AI make possible, and those involved in life in the street.

## 5.2 The designed invisibility of AI in the street

Asymmetric relations between AI-enabling technological infrastructures and situated human actors in the street were further accentuated in observations of the *designed invisibility* of AI. As noted, it has become a truism to say that AI systems are opaque and manifest as a "black box" [8] to scientists and lay audiences alike. However, our AI street observatories made it apparent that this opacity of AI is not only an inherent methodological feature of machine learning-based systems but is designed into the socio-material infrastructures that enable the functioning of AI-based technologies in the street. In looking for "AI" in the street, participants made various attempts to characterise its mode of existence in this environment, describing it not just as 'invisible,' but as 'over their heads' and 'withdrawn,' thus making the association with a feeling of being excluded and ignored by "AI innovation."

During the sensing walk in Coventry, residents and visitors observed various boxes, including wifi routers and "roadside units" high up on lampposts and traffic lights. These boxes form the back-end of the *Midlands Future Mobility* data and communications system that was created to enable automated vehicle testing in this street, but participants in the sensing walk had little to say about them. As one participant noted:

"The cameras in the lampposts, they do not communicate with us, they are above our heads, literally, they communicate with elsewhere […] These boxes are not giving anything, they are just extracting. They seem designed not to draw attention to themselves."

Participants thus located the equipment that enables the functioning of AI in the street outside of the zone of everyday awareness and interaction in the street. In Logan, too, much of the infrastructure for drone delivery was located beyond lines of sight: the so-called drone "nests" that served as operational centres for loading and deployment were located behind barbed wire fences and newly constructed walls, blocking the drones from view. In Edinburgh, participants remarked that having their interactions read through card readers, a case we discussed above, simply did not meet the threshold of relevance in and as everyday life: why care?

While infrastructural invisibility is a familiar trope [38], the placement of AI-enabling technologies outside of the zones of interaction in everyday street settings takes on special significance within the setting of the street: in UK streets today, there are several provisions for technological and data-related transparency and accountability in place today. Indeed, in the Holyhead Road in Coventry, the use of speed cameras is indicated by a traffic sign with a camera symbol, and there is a notice board noting "ANPR in use" - indicating that automated number plate recognition is operating here. There is also a poster notifying the public of the Coventry Council planning application for the installation of road furniture in support of the autonomous vehicle testing in this street. It is thus in the presence of various arrangements of public notification for specific data and monitoring applications, that "AI" is characterised as invisible and "over people' heads" and as leaving people feeling left out of the loop. One Coventry participant responded to technical equipment attached to a lamppost: "I can't take this in as a human, it is not designed for me." In the diagramming session after the listening walk, another commented: "There is all this messaging …, do this, don't do that. But it goes in one direction. Messaging like that is not an interaction, not a negotiation, but a way of giving orders." Here, designed invisibility is read by participants as *a technologically enabled form of exclusion*: participants decode a presence of AI in the street, which is kept out of human sight, in terms of a wilful exclusion of people from active engagements with them.

The sentiment of "AI" in the street being machine-facing rather than human-facing came up across all observatories. In Cambridge, for instance, those on an access data walk came across a hardly noticeable flyer attached to a lamppost and not far from a box disrupting pedestrian traffic flow (Fig. 5a/b). Participants in the walk commented on the need for such a flyer, but such efforts at transparency were seen to further highlight the degree to which AI deployment is detached from the situational use of the street. Through the sensor box on the street in Cambridge, access to the sidewalk was found to be enumerated, and evaluated in numerical terms by and for local government, creating generic metrics of pedestrian counts that could, for example, be used to back new, accessible street designs or the introduction of street furniture. Yet these measurement technologies were seen to override if not elide the actual, situationally contingent ways people must navigate streets—how people in actuality cross streets. In Coventry, this indifference to the situational use of the street by human pedestrians was coded in terms of a handing over of the street environment to the dominant automobile industry. For one participant, the implications felt material: "It can feel like liveable space is confined to a narrow channel of pavement and some small traffic islands. Insofar as there is AI in this street, like the smart traffic light, it operates in this same mode, it is about the cars not people."

There was thus a strong sense in Coventry, as well as in Cambridge, that the introduction of AI in the street involved and aligned with an existing "invisibilisation" of pedestrians and other human road users in this environment. Attempts at transparency, such as traffic signs and notification boards, were seen as inadequate in addressing this re-ordering of space and relations as a consequence of the introduction of smart or intelligent infrastructure. That is also to say, from the standpoint of the street, AI is not necessarily framed or understood as innovative. Instead, AI innovation in several of our observatories was characterised as parasitical, consolidating and amplifying the dominance of and power relations inscribed in existing infrastructures in the street such as automobility (cars) over pedestrians (Coventry) and disadvantaging wheelchair users (Cambridge and Coventry). In Cambridge, the notification of AI-enabled cameras appended to a lamp post was attached at eye level for a standing person and illegible from the sit-point of a wheelchair user. The Holyhead Road was too unsafe to cross in a wheelchair, observed a sister of a wheelchair user. From the vantage point of the street, the invisibility of AI systems and socio-material forms of inaccessibility (illegible notification) merge into more general and sometimes vague narratives regarding the exclusionary experience of technological innovation in the street. The lack of technical specificity of the characterisation of AI in the street by participants, it seems to us, is not only an effect of AI's opacity, or participants' lack of literacy, but also indicates a socio-material indifference inscribed in AI-enabling equipment to human presence in the street.

## 5.3 Statistical governance of the street

Many in participatory design will be familiar with the work of the writer and urban planner, Jane Jacobs, and her now celebrated commentary on city planning in the US in the 1960s. Jacobs critiqued the urban renewal that prevailed in her time. Her well-known remarks on the clearance of tenements spoke to a tiered or stratified model of governing the city. For her, the mechanistic approaches to surveying urban neighbourhoods and governing city planning from above rendered passive those who live on the street. New technologies of city planning meant that citizens "could be dealt with intellectually like grains of sand, or electrons or billiard balls. The larger the number uprooted, the more easily they could be planned for on the basis of mathematical averages" [p. 437, 20]. This approach imposed an abstracted reality onto the city, one in which governing logically happened from above.

> "With statistical and probability techniques [i]t became possible … to map out master plans for the statistical city, and people take these more seriously, for we are all accustomed to believe that maps and reality are necessarily related, or that if they are not, we can make them so by altering reality." [p. 438, 20]

With "eyes on the street" as a mantra, of sorts, Jacobs' call was to reactivate the public in the face of this rendering passive of everyday people, to assert that it is people on the ground and their everyday causal interactions make cities liveable. Her vision was of a governmentality that moves the emphasis away from bureaucratic procedures—like probability techniques —and towards a relational politics, that centres the everyday relations and rhythms of those who make the city through their situated activities.

Our participants' observations of AI in the street evoked Jane Jacobs work for us: the 'invisibility' of AI in the street does not only indicate its close alignment with abstracted modes of governance from afar but also enables a tiered model of governance where the deployment of statistics that renders passive people in the street. Sometimes in positive, but more often in negative terms, the observatories surfaced examples of this statistical mode of governance enabled by AI. On the more positive side, a council representative on the Cambridge data access walk, spoke of AI-based vision recognition and pedestrian counting systems as a way to categorise and "neutralise" particular forms of risk such as obstacles to wheelchairs. In London, observatory participants speculated on rule-based systems to track and respond to the needs of sight-impaired users. Systems might, for instance, intelligently illuminate pathways or change traffic signals depending on need. In Edinburgh, a participant working at a local charity, spoke of the importance of Google's translate app in enabling non-English speaking immigrant populations to extend their networks and find employment.

However, in other instances, statistical modes of governance were perceived as exclusionary, as only affording participation through highly scripted acts (e.g., drone delivery pick up; reviewing the planning application for the automated vehicle testbed) that feel one-way. In Coventry, a Holyhead resident's analogy between the local intelligent vehicle trial and Google Street View bluntly reiterates this point:

"Google Streetview - remember? We had no say in it. One could request one's face to be blurred. They observe but are unobservable themselves."

In Logan, an older couple, who originally used and were excited with the drone delivery system, reflect on the company's reasons for servicing bigger businesses and removing local shops and suppliers from their platform.

**Colleen**: I think the business model changed and I think they were looking for expansion into greater markets. I mean we never ever found out a lot but all of a sudden the small businesses weren't there and Coles appeared….
**Stuart**: They pulled out virtually overnight
Colleen: … and it was so rapid with very limited communication…
**Stuart**: It reeked of a…
**Colleen**: … you know, you've been around business long enough… ok the business model's changed. They want to sign up with the big guns and therefore it pushes the small people out.

Colleen and Stuart speculate on the drone delivery company's motives, as driven by a commercial logic prioritising logistical optimisation and expansion; a model where 'the small people' help to test and expand an algorithmic reasoning and are then replaced by the 'big guns.' Right or wrong, the net result is a commercially managed technical tier that governs access to services. The decisions and the sphere of actors brought into play that shape the decisions are not visible or accountable to the street but have a significant impact on the life lived on it.

In our participants' observations of how AI-enabled technologies operate in the street, then, there are clear routes to a return of a 'statistical city' although the statistical governmentality that assisted governing the city in Jacob's time today seems to have shifted from central government to privately managed, "exstituted" [19] algorithmic systems. On the street, our observatories revealed the danger that public institutions are being perceived as handing over public environments to private industries and of off-loading responsibility for public life (traffic management, safety) to closed or obscured infrastructures of algorithmic reason. This reasoning operates in a dispersed manner across varied, obscured organisations with publics being left to wonder.

## 6 Discussion: The reciprocity deficits of AI in the street

Taken together, the observations and insights that emerged from our street-level observatories of AI point towards a shared problematic: that the introduction and presence of AI in the city streets under scrutiny here suffer from what we term a "reciprocity deficit", a disruption of the responsive relations between actors that organise encounters in city streets and bind a city together. Here we introduce this term and discuss its relevance for our understanding of the street as a space of public engagement with AI.

The importance of reciprocity in city life is highlighted by Jacobs' phrase, "eyes on the street", which places emphasis on a mutual seeing, a joint, situated form of observation that emerges between those living and working together on the street, through their ordinary interactions. Engagements through our observatories indicated that this form of seeing is frequently perceived to be absent or even actively disabled by the data and computational infrastructures that enable AI in the street

It is far from easy to 'see' or build a relation with the material or structural features of AI. There is a type of social invisibility activated by AI in the street that is different from the methodological invisibility targeted in *explainable AI* research, where the end user may not have the tools or capabilities to see what an AI system is doing—the AI's operations are opaque or invisible. However, our participants equally articulated a different type of invisibility in relation to AI, one that arises from (not) being meant to see and not 'being seen' by "AI" in the street as a lived environment.

Here, the operations of AI in the street, which involve large volumes of data and demand sophisticated computational processes to manage and administer it are perceived as not designed for human consumption; they are in their very nature withdrawn from and indifferent to the subjects they "observe and enumerate" [p. 256, 12], or that inhabit the lived environment alongside them. In Cambridge, for example, the situational relevance of data infrastructures only became visible through failures in the built environment and in particular problems with street-level access. Other than that, data and computational logics operated at a bureaucratic, governmental level, remaining out of sight and indifferent to the situational logics of moving through the street. In Coventry, this detachment was felt acutely. As noted (Section 5.2), the automated driving technologies, connected to instrumented lamp posts or controlling traffic lights, were seen as enablers for cars not citizens. Participants perceived these technologies as designed to ignore them, to be mutually invisible in relation to the human. Our first point, then, is to suggest that the invisibility of AI in the street does not only pertain to epistemic logics of transparency but to social logics of the ordering of relations. The invisibility of AI in the street was articulated by participants as a mode of stratification, where the technology and its connections operate above or beyond and with indifference to a human scale.

Extending this idea of seeing and a detachment from the street, we want to return the importance Jacobs attributed to forms of place-based sociality. Jacobs made much of the web of casual interactions that sustain collective respect and trust in city neighbourhoods. We want to argue that, as data, computation and AI pervade public spaces (such as the street), an accompanying *reciprocity deficit* risks becoming complicit in the erosion of this web. The themes that emerged from our everyday AI observatories point to rising tensions, asymmetries and fractures because of—to borrow Jacob's language—a diminished sense of corresponding gaze from equivalent eyes on the street and a reduced mutually accomplished sense of what counts as safe and trustworthy. Through mundane systems such as loyalty card programmes and innovative commercial infrastructures that enable for instance food delivery by drone, there is a feeling of being observed by others, but in a way that is not 'on the street' or 'for the street'—that is not reciprocal.

In our participants' observations of reciprocity deficits, descriptions of the technological features of AI-based systems and of the politico-economic models underpinning AI innovation in the city were closely interwoven. The remotely monitored advertising board in Edinburgh, or the trials of intelligent vehicle and drone delivery APIs in Coventry and Logan were perceived as operating at scales and in markets that are detached from the situated environment of the street and of little direct benefit to local communities. "The advertising", Ned asserts, "has nothing to do with Edinburgh", the drone trial in Logan, Michael laments, "doesn't benefit… people that are living here".

People's views from the street, together with the thinking from Jacobs, then help us to project a wider commentary on the pervasive presence of AI in the observed streets. Just as Jacobs' critique linked changes to the built environment to new yet troubling models of the social, we argue that data, computation and AI on the street are figuring different structural relations; the attendant gulf between those on the street and those managing the street seem to be eroding the features of street life that allow for mutuality in trust and responsibility. This is what we're referring to as the reciprocal deficit in AI on the street. Rather than a web of sociality that creates, re-iterates or affirms alliances, this deficit seems to contribute to a sense of mistrust in public spaces. The sense is of a public realm where the cycle of give-and-take and back-and-forth is ruptured or disentangled. Adie, in trying to capture this, talks of a disruption to the "continuity of things" on Leith Walk in Edinburgh. Evoking Augé's [2] *non-places* he pronounces: "I call it the gamma land - the grey space."

Also striking about our observatories, however, was an apparent refusal to linger on this discontent. During our observatory activities, we found many participants looking for a more productive respecification of AI, expressing a drive to co-exist constructively with AI and engage with its transformative potential; not to accept the infrastructural and social invisibility of AI and the covering over of a failure to innovate, but to express a curiosity and commitment to making AI work on different terms. In Logan, the promise was felt to be in enabling local commerce and, in Edinburgh, enabling precarious communities. In Cambridge, Coventry and London the technology-infused environment and sensing technologies were discussed not just in extractive terms, but for their potentiality to enable sustainable mobility, refigure the relations between public and governing bodies, transforming participation in planning, pursuing disability justice, and articulating rights to movement and safety.

# 7 Conclusion

Our *in-situ* design interventions, in the form of sensing walks and diagramming workshops, then enabled multiple, contrasting, articulations of AI in the street, which enables us as researchers to specify its problematic in place-based or situational terms: on the one hand, the introduction of AI-based applications into street environments was found to display a reciprocity deficit. People expressed a sense of exclusion - of being "unseen," left out of the loop and ignored - by means of algorithmic governance, even as they are being traced, counted and/or monitored by smart computational infrastructures in the street. At the same time, the equipment of the street as an AI innovation environment was perceived by many participants in our observatories as an opportunity to redefine relations between the technological economy and people in the street, and many did not hesitate to allocate capacities for social problem-solving to AI.

These findings demonstrate the relevance of Jane Jacobs analysis of urban planning to AI innovation in cities. To understand the 'impact' of AI on urban communities, it is crucial that we observe it in relation to and in correspondence with the ways people live in and use living environments of the street. The credibility and viability of AI as a solution for shared problems demands that decision makers not only "understand specific services and techniques…" but grasp the necessity to "understand, and understand thoroughly, specific places" [p. 410, 20]. To address the problems of AI in society, we must move beyond making the technicities of AI explainable. We must consider how AI in the street intervenes in and operates on the cycles of everyday giving and receiving that are bound to specific places and that create and sustain the essential ingredients of what Jacobs terms "vital streets."

Our paper also makes a methodological contribution. Our everyday AI observatories provide a way to activate a situational orientation to AI in society, enabling an articulation of technological infrastructures from the standpoint of its operations on the situated relations and unfolding contingencies of life in the street. This kind of situational street was in fact what Jacobs conjured up in her commentary and through her use of "eyes on the street". Jacobs' was not a mechanistic seeing, but a model of the social that depicted the situated, casual and responsive interactions on the street as essential to its sustenance and ordering, to the forms of respect, liveliness, responsibility and safety that sustain life on the street. Crucial to the "relational analytic" that was activated and enabled by our everyday AI observatories was their double function: our interventions enabled at once engagement with and collaborative inquiry into AI in the street. This connection between participation in a common world and its exploration has long been recognised as key to participatory design research [21, 28]. In our work we infused this approach with conceptual insights from social and cultural studies of AI, STS and urban studies. Our aim in doing so is to strengthen the capacity of design methodologies to respecify the social and public problem of technology & community in our computationally intensive, ecologically challenged society.

This paper then offers the everyday AI observatory as a way to engage people in the specification of situational requirements on AI in lived environments of the street. Staged in a primary site where everyday publics encounter AI, this method provides a means to invite public commentary on and participation in specifying the role of AI in urban environments. The trial of five observatories (in Cambridge, Coventry, Edinburgh, Logan and London) and placing their observations and insights in conversation with related scholarship in public and material participation in technological societies, foregrounds the importance of attending to how AI operates within and upon situational contingencies in the street. A priority that follows from this is to define the conditions in which greater reciprocity between everyday publics and AI/data infrastructures in the city would offer a credible path to addressing problems of exclusion in urban settings and expanding social agency within lived environments.

## ACKNOWLEDGMENTS

We thank all the participants in the everyday AI observatories that we conducted in Cambridge, Coventry, Edinburgh, London and Logan (Australia) in the summer of 2024, as well as to our project partners and funders.